\documentclass[epj,referee]{svjour}
\usepackage{latexsym}
\usepackage{amssymb}
\usepackage{amsmath}
\usepackage{mathrsfs}
\usepackage{graphics}
\usepackage{color}

\renewcommand{~}{\,}

\newcommand{\R}{\mathbb{R}}
\newcommand{\NV}{\Delta_{\mathtt{N}}}
\newcommand{\TV}{\Delta_{\mathtt{T}}}

\newcommand{\e}{\mathtt{e}}
\newcommand{\p}{\wp}

\newcommand{\dr}{\mathtt{d}r}
\newcommand{\ds}{\mathtt{d}s}
\newcommand{\A}{\mathcal{A}}
\newcommand{\K}{\mathcal{K}}
\newcommand{\N}{\mathcal{N}}
\renewcommand{\rho}{\varrho}
\newcommand{\Tbar}{{\overline{T}}}
\newcommand{\nbar}{{\overline{n}}}

\renewcommand{\leq}{\leqslant}

\newcommand{\BE}{\begin{equation}}
\newcommand{\EE}{\end{equation}}

\begin{document}
\definecolor{cihlova}{rgb}{0.7969,0.1992,0}
\title{Spectral rigidity of vehicular streams (Random Matrix Theory approach)}
\author{Milan Krb\'alek\inst{1}\inst{2} and Petr \v
Seba\inst{2}\inst{3}\inst{4}}
\institute{Faculty of Nuclear Sciences and Physical Engineering,
Czech Technical University in Prague, Prague - Czech Republic,
\and Doppler Institute for Mathematical Physics and Applied
Mathematics, Faculty of Nuclear Sciences and Physical Engineering,
Czech Technical University in Prague, Prague - Czech Republic\and
University of Hradec Kr\'alov\'e, Hradec Kr\'alov\'e - Czech
Republic\and Institute of Physics, Academy of Sciences of the
Czech Republic, Prague - Czech Republic\\Corresponding author
e-mail:~\texttt{\textcolor{blue}{milan.krbalek@fjfi.cvut.cz}} }
\date{Received: date / Revised version: date}
%
\abstract{Using the methods originally developed for Random Matrix
Theory we derive an exact mathematical formula for number variance
$\NV(L)$ (introduced in \cite{Red_cars}) describing a rigidity of particle ensembles with power-law repulsion. The
resulting relation is consequently compared with the relevant statistics
of the single-vehicle data measured on the Dutch freeway A9. The
detected value of an inverse temperature $\beta,$ which can be
identified as a coefficient of a mental strain of car drivers, is
then discussed in detail with the respect to the traffic density
$\rho$ and flow $J.$
} 
\maketitle


\section{Terminus a quo}

The main goal of this paper is to show that the statistical
fluctuations of single-vehicle data (in vehicular flows) can be
very well predicted by the methods known from Random Matrix Theory
where used for describing the statistics of energy levels in
quantum chaotic systems. Above that, we intend to demonstrate that
the changes of statistical variances in vehicular samples depends
not only on macroscopical traffic quantities and three traffic
phases (as published in \cite{Kerner} or \cite{Helbing}), but also
on psychological characteristics of driver's decision-making
process. We will evince that transitions among the traffic
phases (free flow, synchronized flow, and wide moving jam) cause
perceptible changes in the mental strain of car drivers.\\

As reported in Ref. \cite{Kerner}, \cite{Helbing}, \cite{Knospe},
\cite{Red_cars} and \cite{Kerner2} three traffic phases show
substantially different microscopical properties. It was
demonstrated in \cite{Helbing_and_Krbalek}, \cite{Gas},
\cite{Wilson}, and \cite{HKT} that the microscopical traffic
structure can be estimated with the help of a repulsive potential
(applied locally only) describing the mutual interaction between
successive cars in the chain of vehicles. Especially, the
probability density $\p(r)$ for the distance $r$ of two
subsequent cars (\emph{clearance distribution}) can be derived by
means of an one-dimensional gas having an inverse temperature
$\beta$ and interacting by the repulsive potential $V(r)=r^{-1}$ (as
discussed in Ref. \cite{Gas}, \cite{Helbing_and_Krbalek}, and
\cite{Wilson}). Concretely, denoting $\Theta(x)$ the Heaviside's
function
$$\Theta(x)=\left\{\begin{array}{ccc} 1,&\hspace{0.2cm}&x>0\\0,&\hspace{0.2cm} & x\leq
0,\end{array}\right.$$
and $\K_\lambda(x)$ the modified Bessel's function of the second
kind of order $\lambda$ (Mac-Donald's function), the clearance
distribution of the above-mentioned thermodynamical traffic gas reads as
\BE \p(r)=\A~ \Theta(r)~\e^{-\frac{\beta}{r}}\e^{-Br},
\label{p_beta} \EE
where \BE B=\beta+\frac{3-\e^{-\sqrt{\beta}}}{2}, \label{becko}
\EE
\BE
\A^{-1}=2\sqrt{\frac{\beta}{B}}\K_1\bigl(2\sqrt{B\beta}\bigr).\label{acko}
\EE
We remark that $\p(r)$ fulfils two normalization conditions
\BE \int_\R \p(r)~\dr=1 \label{norma1} \EE
and
\BE \int_\R r~\p(r)~\dr=1. \label{norma2} \EE
The latter represents a scaling to the mean clearance equal to
one, which is introduced for convenience. The above-mentioned
distribution (\ref{p_beta}) is in a good agreement with the
clearance distribution of real-road data (\cite{Helbing_and_Krbalek}, \cite{Gas}, \cite{Red_cars}, and
\cite{Wilson}) whereas the inverse temperature $\beta$  is related to the traffic density $\rho$. We note that the inverse temperature $\beta$ can be understood as a quantitative
description of the mental strain of drivers in a given traffic situation. More specially, the parameter $\beta$ reflects the psychological pressure level under which the driver is if moving in the traffic stream. Whereas the free flows induces practically no mental strain of drivers, with the increasing traffic density the mental strain escalates rapidly. This will be confirmed in the following text.

\section{Statistical variances in traffic data}

Another powerful way to inspect the interactions between cars within the
highway data is to investigate the traffic flow fluctuations. One
possibility is to use  the so-called \emph{time-gap variance} $\TV$
considered in paper \cite{Dirk_and_Martin} and defined as follows.
Let $\{t_i:i=1 \ldots Q\}$ be the data set of time intervals between
subsequent cars passing a fixed point on the highway. Using it one
can calculate the moving average
$$T_k^{(N)}=\frac{1}{N} \sum_{i=k}^{k+N-1} t_i \hspace{0.5cm} (k=1\ldots Q-N+1)$$
of the time intervals produced by the $N+1$ successive vehicles
(i.e. $N$ gaps)
as well as the total average
$$\Tbar=\frac{1}{Q}\sum_{i=1}^Q t_i\equiv T_1^{(Q)}.$$
The time-gap variance $\TV$ is defined by the variance of the
sample-averaged time intervals as a function of the sampling size
$N,$
$$\TV=\frac{1}{Q-N+1}\sum_{k=1}^{Q-N+1} \left(T_k^{(N)}-\Tbar\right)^2,$$
where $k$ runs over all possible samples of $N+1$ successive cars.
For time intervals $t_i$ being statistically independent the law
of large numbers gives $\TV(N)\propto 1/N$.\\

A statistical analysis of the data set recorded on the Dutch freeway
A9 and published in Ref. \cite{Dirk_and_Martin} leads, however, to
different results - see the Figure 1. For the free traffic flow
$(\rho< 15~\mathrm{veh/km/lane})$ one observes indeed the
expected behavior $\TV(N)\propto 1/N$. More interesting behavior,
nevertheless, is detected for higher densities $(\rho> 35~
\mathrm{veh/km/lane}).$ Here Nishinari, Treiber, and Helbing (in
Ref. \cite{Dirk_and_Martin}) have empirically found a power law
dependence
$$\TV(N)\propto N^\epsilon$$ with an exponent $\epsilon \approx
-2/3,$ which can be explained as a manifestation of correlations
between the queued vehicles in a congested traffic flow.\\

\begin{figure} \centering
\scalebox{.4}{\includegraphics{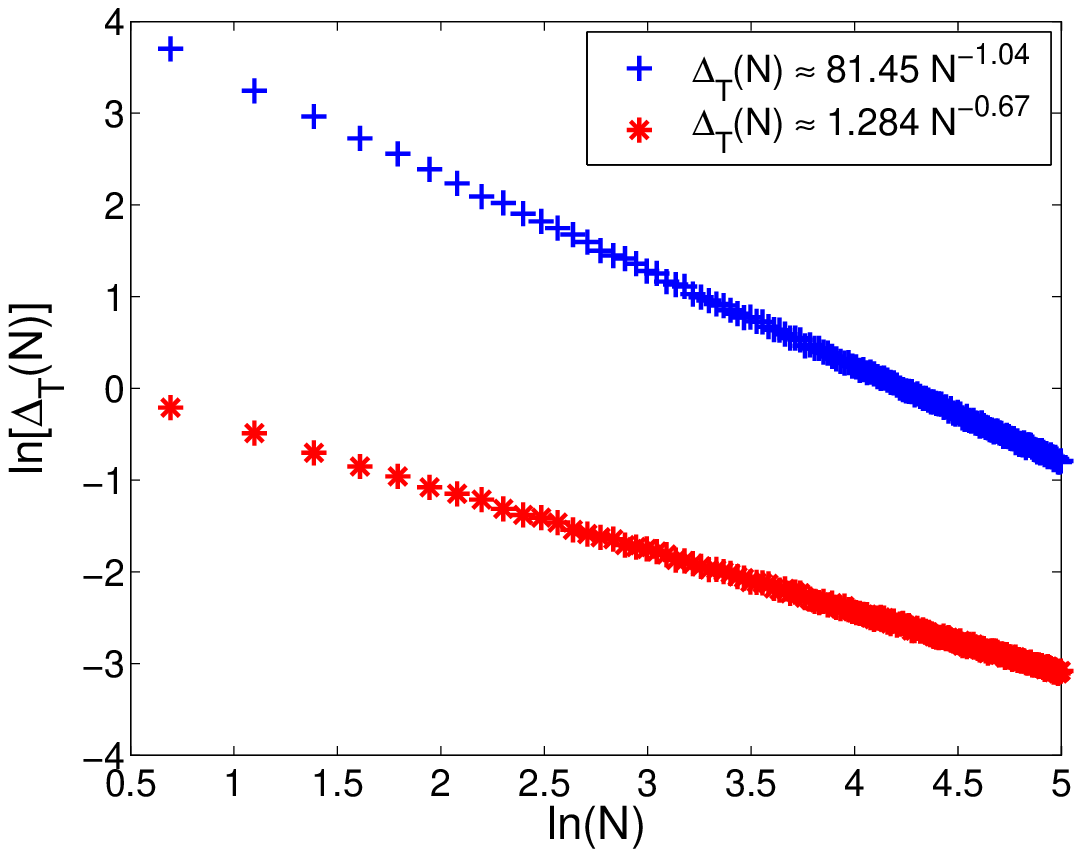}} {\caption{
\small The time-gap variance $\TV(N)$ as a function of the sampling
size $N$ (in log-log scale). Plus signs and stars represent the
variance of average time-gaps for free and congested flows,
respectively.}} \label{TimeNV}
\end{figure}

There is, however, one substantial drawback of this description. The
time-gap variance was introduced \emph{ad hoc} and hardly anything
is known about its exact mathematical properties in the case of
interacting vehicles. It is therefore appropriate to look for an
alternative that is mathematically well understood.  A natural
candidate is the \emph{number variance} $\NV(L)$ that was originally
introduced for describing \emph{a spectral rigidity} of energy
levels in quantum chaotic systems, i.e. for describing a structure
of eigenvalues in the Random Matrix Theory. $\NV(L)$ reproduces also
the variances in the particle positions of a certain class of
one-dimensional interacting gases (for example Dyson gas in Ref.
\cite{Mehta}) and
it is defined as follows.\\

Consider a set $\{r_i:i=1 \ldots Q\}$ of  distances (i.e.
\emph{clearances} in the traffic terminology) between each pair of
subsequent cars moving in the same lane. We suppose that the mean distance
taken over the complete set is re-scaled to one, i.e.
$$\sum_{i=1}^Q r_i=Q.$$
Dividing the interval $[0,Q]$ into subintervals $[(k-1)L,kL]$  of
a length $L$ and denoting by $n_k(L)$ the number of cars in the
$k$th subinterval, the average value $\nbar(L)$ taken over all
possible subintervals is
$$\nbar(L)=\frac{1}{\lfloor Q/L \rfloor} \sum_{k=1}^{\lfloor Q/L
\rfloor} n_k(L)=L,$$
where the integer part $\lfloor Q/L \rfloor$ stands for the number
of all subintervals $[(k-1)L,kL]$ included in the interval $[0,Q].$
Number variance $\NV(L)$ is then defined as
$$\NV(L)=\frac{1}{\lfloor Q/L \rfloor} \sum_{k=1}^{\lfloor Q/L
\rfloor} \left(n_k(L)-L\right)^2$$
and represents the statistical variance of the number of vehicles
moving at the same time inside a fixed part of the road of a length
$L.$ The mathematical properties of the number variance are well
understood and therefore $\NV(L)$ serves as a better alternative for
description of traffic fluctuations (a rigidity of the vehicular
chain) than the time-gap variance $\TV(N)$ itself (see also \cite{Red_cars}).

\section{Exact formula for number variance}

Considering the probability density (\ref{p_beta}) with the only
free parameter $\beta$ (the inverse temperature) we aim to derive an
exact formula for the rigidity $\NV(L)$ of thermodynamical traffic
gas (see \cite{Gas}). For these purposes we use the methods
presented in \cite{Mehta} and \cite{Bogomolny} in
detail.\\

\begin{figure} \centering
\scalebox{.4}{\includegraphics{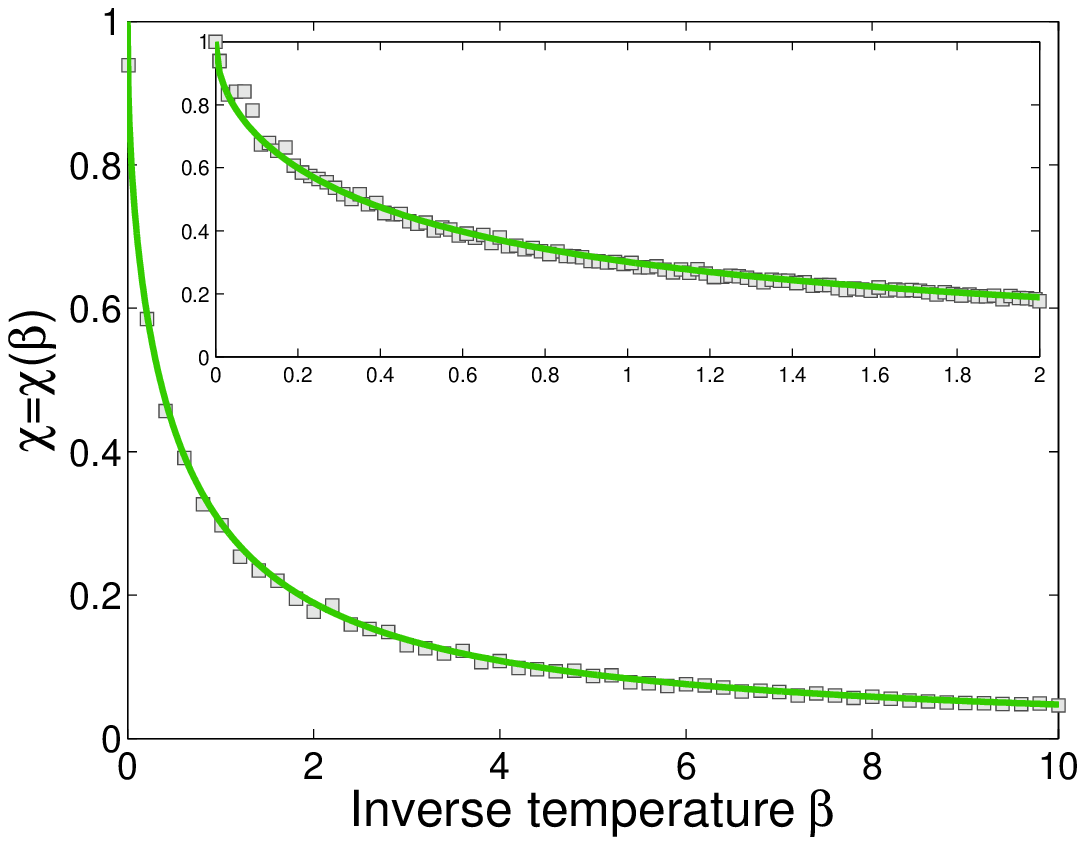}}\\
{\caption{ \small The slope $\chi(\beta)$ of the number variance $\NV(L)=\chi L +\gamma $ as a function of the inverse temperature $\beta.$ The squares represent the results of numerical calculations whereas the continuous curve displays the result (\ref{forestGUMP_chi}) obtained by the exact mathematical computations. The behavior close to the origin is magnified in the inset.}} \label{ExactNV_chi}
\end{figure}

Let $\p_n(r)$ represent the $n$-th nearest-neighbor probability
density ($n$-th probability density for short), i.e. the
probability density for the spacing $r$ of $n+2$ neighboring
particles. Owing to the fact that probability density for spacing
between two succeeding particles (cars) is $\p(r)=\p_0(r),$ the
$n$-th probability density $\p_n(r)$  can be calculated via
recurrent formula
$$\p_n(r)=\p_{n-1}(r)~\star~\p_0(r),$$
where symbol $\star$ represents a convolution of two independent probabilities, i.e.
$$\p_n(r)=\int_{\R} \p_{n-1}(s)\p_0(r-s)~\ds.$$
Using the approximation of the function
$$g_n(s)=\e^{-\beta\bigl(\frac{n^2}{s}+\frac{1}{r-s}\bigr)}$$ in the saddle point one can obtain
$$\p_n(r)=\Theta(r)\N_nr^n\e^{-\frac{\beta}{r}(n+1)^2}~\e^{-Br},$$
where
$$\N_n^{-1}=2\left(\sqrt{\frac{\beta}{B}}(n+1)\right)^{n+1}\K_{n+1}\bigl(2(n+1)\sqrt{B\beta}\bigr)$$
fixes up the proper normalization $\int_\R \p_n(r)~\dr=1.$ In
addition to that the mean spacing equals $$\int_\R r\p_n(r)~\dr=n+1.$$
According the book \cite{Mehta} the variance $\NV(L)$ could be
evaluated by the formula
\BE \NV(L)=L-2\int_0^L (L-r)\bigl(1-R(r)\bigr)~\dr, \label{hlavni}
\EE
where
$$R(r)=\sum_{n=0}^\infty \p_n(r)$$
is the \emph{two-point cluster function.} The convenient way to
calculate the asymptotic behavior of the variance $\NV(L)$ for large
$L$ is the application of the Laplace transformation to the two-point cluster
function $y(t)=\int_\R R(r)~\e^{-rt}~\dr.$ It leads to a partial
result
$$y(t)=\sum_{n=0}^\infty
\left(\frac{B}{B+t}\right)^{\frac{n+1}{2}}\frac{\K_{n+1}\bigl(2(n+1)\sqrt{(B+t)\beta}\bigr)}{\K_{n+1}\bigl(2(n+1)\sqrt{B\beta}\bigr)}.$$
The small$-x$ asymptotic behavior of the Mac-Donald's function
$$\K_n(x) \approx \frac{2^{n-1}\Gamma(n)}{x^n}~\e^{-x} \hspace{1cm} (x \ll 1)$$
(where $\Gamma(x)$ represents the gamma-function) provides the approximation
$$y(t)=\left(\frac{B+t}{B}~\frac{\e^{2\sqrt{(B+t)\beta}}}{\e^{2\sqrt{B\beta}}}-1\right)^{-1}.$$
Applying the Maclaurin's expansion (Taylor's expansion about the
point zero) of the function $h(t)=t \cdot y(t)$ to order $t^2$ we
obtain
$$y(t)\approx\frac{1}{t}+\alpha_0+\alpha_1t+\mathcal{O}(t^2),$$
where
$$\alpha_0=-\frac{2B\beta+3\sqrt{B\beta}}{4\bigl(1+\sqrt{B\beta}\bigr)^2},$$
$$\alpha_1=\frac{6\sqrt{B\beta}+B\beta\bigl(21+4B\beta+16\sqrt{B\beta}\bigr)}{48B\bigl(1+2\sqrt{B\beta}\bigr)^3}.$$
Then we get from equation (\ref{hlavni})
\BE \NV(L)=\chi L +\gamma + \mathcal{O}(L^{-1}),\label{forestGUMP}
\EE
where
\BE \chi=\chi(\beta)=\frac{2+\sqrt{B\beta}}{2B(1+\sqrt{B\beta})}\label{forestGUMP_chi}
\EE
and
\BE \gamma=\gamma(\beta)=\frac{6\sqrt{B\beta}+B\beta\bigl(21+4B\beta+16\sqrt{B\beta}\bigr)}{24\bigl(1+\sqrt{B\beta}\bigr)^4}.\label{forestGUMP_gamma}
\EE
This crowns the effort to derive an exact form for the number
variance. The linear dependence (\ref{forestGUMP}) represents a
large$-L$ approximation and its correctness is demonstrated in the
Fig. 2 and Fig. 3 where compared to the results of numerical computations.

\begin{figure} \centering
\scalebox{.4}{\includegraphics{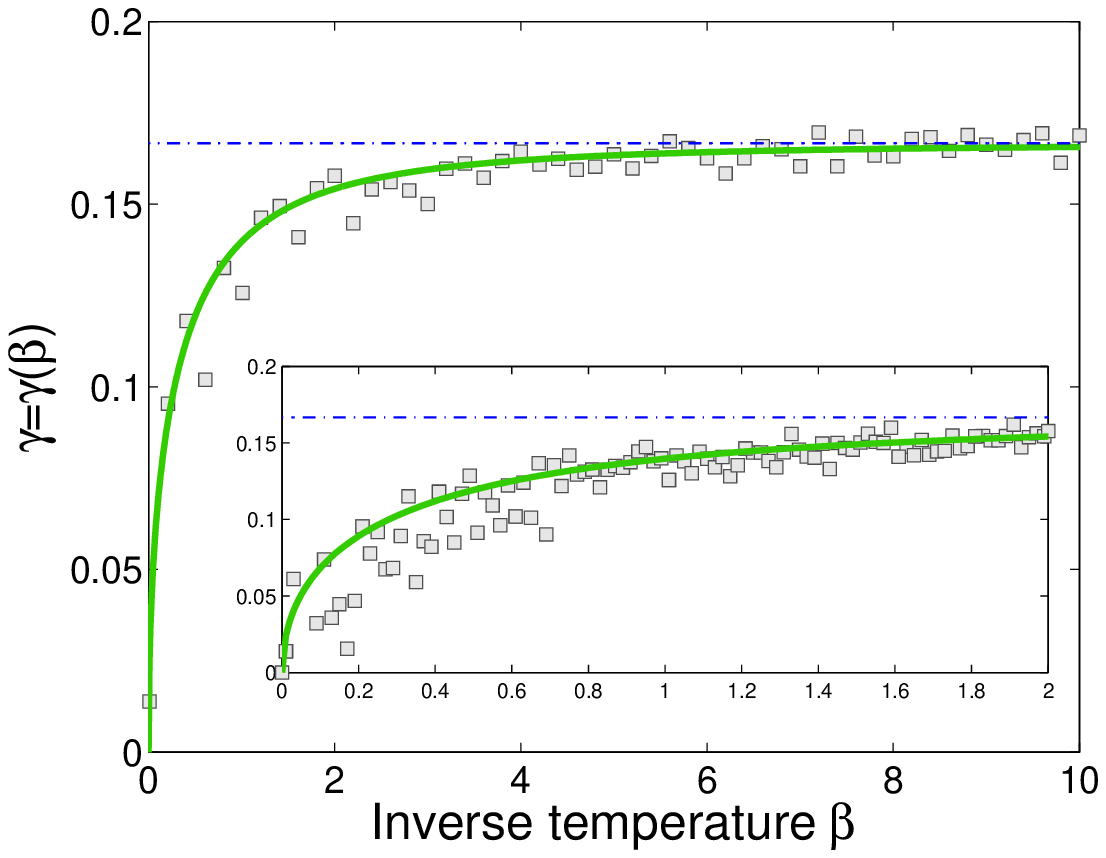}}\\
{\caption{ \small The shift $\gamma(\beta)$ of the number variance $\NV(L)=\chi L +\gamma $ as a function of the inverse temperature $\beta.$ The squares represent the results of numerical calculations whereas the continuous curve visualizes the result (\ref{forestGUMP_gamma}) obtained by the exact mathematical computations. The behavior close to the origin is magnified in the inset. The dash-dotted line displays the asymptotic tendency in $\gamma=\gamma(\beta),$ i.e. $\lim_{\beta \rightarrow \infty} \gamma(\beta)=1/6.$
 }} \label{ExactNV_gamma}
\end{figure}

\section{Number variance of traffic data }

As already discussed the behavior of the number
variance is sensitive to the temperature $\beta$ -
or in the terminology of the Random Matrix Theory - to the
universality class of the random matrix ensemble. To use the known
mathematical results one has not to mix together states with
different densities - a procedure known as \emph{data unfolding}
in the Random Matrix Theory. For the transportation this means
than one cannot mix together traffic states with different traffic
densities (as published in \cite{Dirk_and_Martin}) and hence with a different vigilance of the drivers. So we will perform a separate analysis of the data-samples lying within
short density intervals to prevent so the undesirable mixing of
the different states.\\

\begin{figure} \centering
\scalebox{.4}{\includegraphics{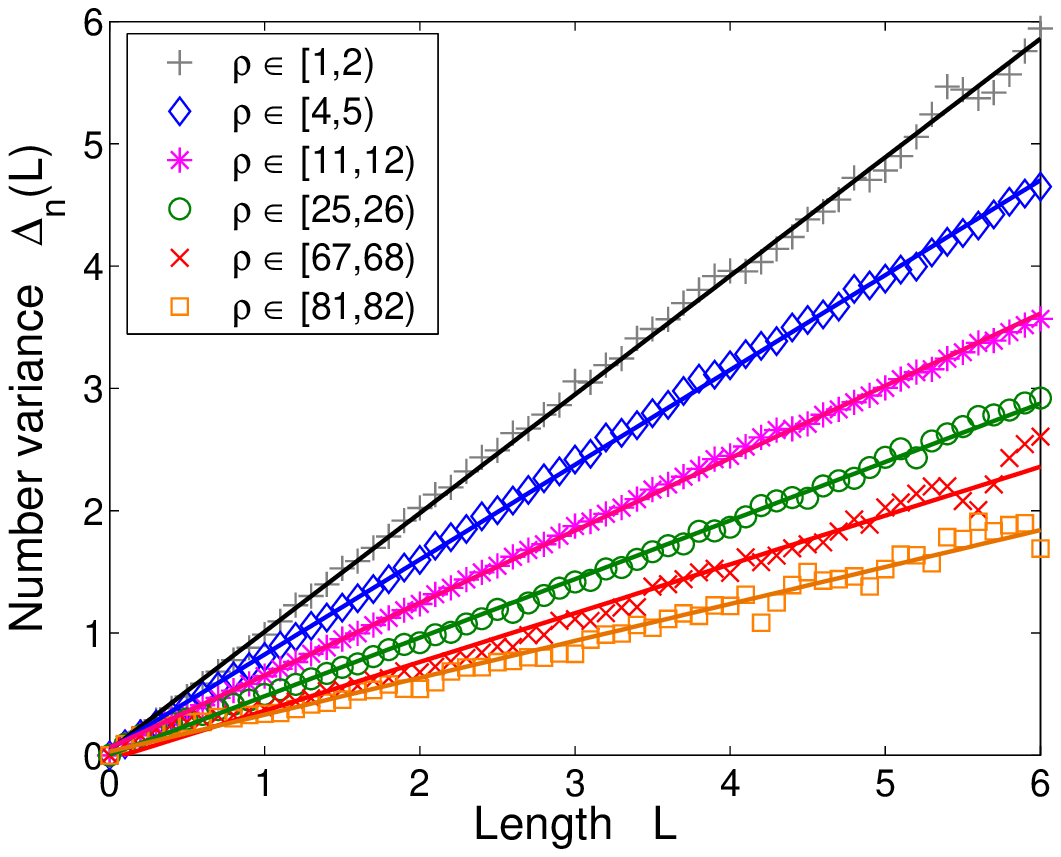}}
\caption{ \small The number variance $\NV(L)$ as a function of
the length $L.$ Plus signs, diamonds, stars, circles, crosses, and
squares represent the number variance of real-road data in the
selected density regions (see the legend for details). The curves show
the linear approximations of the individual data. Their slopes
were carefully analyzed and consecutively visualized in the Fig. 5
(top part).}
\end{figure}

Anyway, we divide the region of the measured densities $\rho \in
[0,85~\mathrm{veh/km/lane}]$ into eighty five equidistant
subintervals and  analyze the data from each one of them
separately. The number variance $\NV(L)$ evaluated with the
data in a fixed density interval has a characteristic linear tail
(see Fig. 4) that is well known from the Random Matrix Theory.
Similarly, such a behavior was found in models of one-dimensional
thermodynamical gases with the nearest-neighbor repulsion among
the particles (see Ref. \cite{Bogomolny}). We remind that for the
case where the interaction is not restricted to the nearest
neighbors but includes all particles the number variance has
typically a logarithmical tail - see \cite{Mehta}. So the linear
tail of $\NV(L)$  supports the view that in the traffic
stream the interactions are restricted to the few nearest cars
only. The slope of the linear tail of  $\NV(L)$ decreases
with the traffic density (see the top subplot in the Fig. 5). It
is a consequence of the increasing alertness of the drivers and
hence of the increasing coupling between the neighboring cars in
the dense traffic flows.\\

\begin{figure} \centering
\scalebox{.4}{\includegraphics{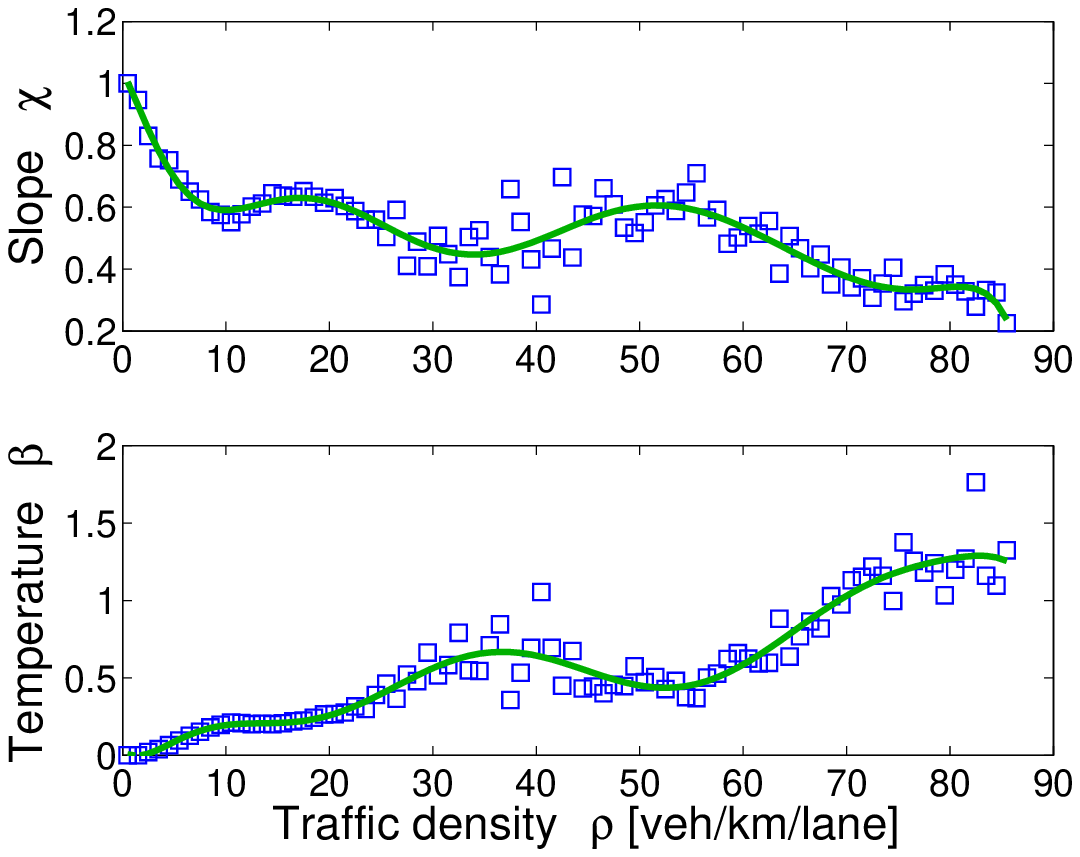}}
\caption{ \small The slope $\chi$ and the inverse temperature
$\beta$ as functions of the traffic density $\rho.$ The squares
on the upper subplot display the slope of the number variance
$\NV(L)$ (see Fig. 4), separately analyzed for various
traffic densities. The lower subplot visualizes the fitted values
of the inverse temperature $\beta,$ for which the exact form of
number variance $\NV(L)=\chi(\beta)~L+\gamma(\beta)$
corresponds to the number variance obtained from the traffic data.
The continuous curves represent polynomial approximations of the
relevant data.}
\end{figure}

The exactly derived properties of the function $\NV(L)$ agree with
the behavior of the number variance extracted from the traffic
data (see Fig. 4). A comparison between traffic data number
variance and the formula (\ref{forestGUMP}) allows us to determine
the empirical dependence of inverse temperature $\beta$ on traffic
density $\rho$. The inverse temperature reflects the
microscopic status in which the individual vehicular interactions
influence the traffic.  Conversely, in the macroscopic approach,
traffic is treated as a continuum and modelled by aggregated,
fluid-like quantities, such as density and flow (see \cite{Kerner}
and \cite{Helbing}). Its most prominent result is the dependence of
the
traffic flow on the traffic density - the fundamental diagram.\\

It is clear that the macroscopic traffic characteristics are
determined by its microscopic status. Consequently there should be
a relation between the behavior of the fundamental diagram and
that of the inverse temperature $\beta$. In the Figure 6 we
display the behavior of  the inverse temperature $\beta$
simultaneously with the fundamental diagram. The both curves show
a virtually linear increase in the region of  a free traffic (up
to $\rho \approx 10~\mathrm{veh/km/lane}$). The inverse
temperature $\beta$ then displays a plateau for the densities up
to $18~\mathrm{veh/km/lane}$ while the flow continues to increase.
A detailed inspection uncovers, however, that the increase of the
traffic flow ceases to be linear and becomes concave at that
region. So the flow is reduced with respect to the outcome
expected for a linear behavior - a manifestation of the onset of
the phenomenon that finally leads to a congested traffic. For
larger densities the temperature $\beta$ increases up to $\rho
\gtrapprox 32~\mathrm{veh/km/lane}$. The center of this interval
is localized at $\rho \approx 25 $ -- a critical point of the
fundamental diagram at which the flow starts to decrease. This
behavior of the inverse temperature is understandable and imposed
by the fact that the drivers, moving quite fast in a relatively
dense traffic flow, have to synchronize their driving with the
preceding car (a strong interaction) and are therefore under a
considerable psychological pressure. After the transition from the
free to a congested traffic regime (between 40 and
$50~\mathrm{veh/km/lane}$), the synchronization continues to
decline because of the decrease in the mean velocity leading to
decreasing $\beta$. Finally - for densities related to the
congested traffic - the inverse temperature increases while the
flow remains constant. The comparison between the traffic flow and
the inverse temperature is even more illustrative when the changes
of the flow are taken into account. Therefore we evaluate the
derivative of the flow
$$J'=\frac{\partial J}{\partial \rho}.$$
The result of the evaluation can be seen from the Figure 5 where one can trace the significant similarities between the shape of $J'=J'(\rho)$ and inverse temperature $\beta.$

\begin{figure} \centering
\scalebox{.4}{\includegraphics{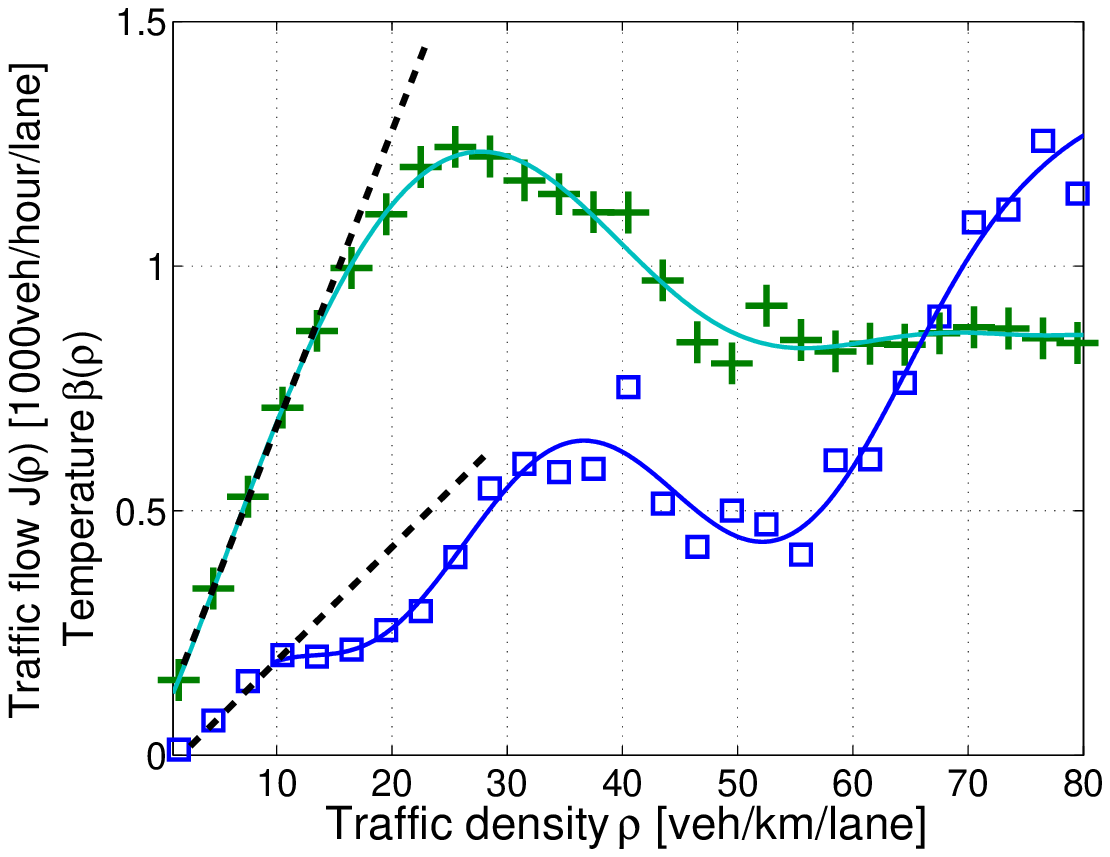}}
\caption{ \small Traffic flow $J(\rho)$ and inverse temperature
$\beta(\rho)$ as functions of the traffic density $\rho$. Plus
signs display a traffic flow in thousands of vehicles per hour and
squares correspond to inverse temperature of the traffic gas. The
results of a polynomial curve-fitting are visualized by the
continuous curves. The dashed lines represent a linear
approximations of the relevant data near the origin.}
\end{figure}

\section{Summary and conclusion}

With the help of methods of Random Matrix Theory we have derived the exact formula for the rigidity (quantified by the number variance $\NV(L)$) of the one-dimensional thermal particle-gas with the repulsion-potential which depends on the reciprocal value of clearances among the nearest-neighboring particles. The correctness of the result obtained was confirmed by the comparison to the numerical solution.\\

Analogously to the investigations presented in \cite{Red_cars} we have analyzed the number variance $\NV$ of single-vehicle data measured on the Dutch freeway A9. The robust statistical analysis (based on the principle of data unfolding known from Random Matrix Theory) revealed that the rigidity of traffic samples (evaluated by means of the number variance $\NV$) shows a linear dependence \BE \NV(L)\approx \chi L+\gamma\label{final}\EE  (in each of eighty five equidistant density-subregions) whose slope $\chi$ depends on traffic density significantly. This knowledge substantially specifies the general results published in Ref. \cite{Dirk_and_Martin}. \\

Above that, the comparison of the exact result (\ref{final}) with the number variance of traffic data has provided the remarkable possibility how to detect the mental strain level of car drivers moving in the traffic stream. It was demonstrated that the inverse temperature of the traffic sample, indicating the degree of stress of the drivers, shows an increase at both the low and high densities. In the intermediate region, where the free flow regime converts to the congested traffic, it displays more complex behavior.\\

To conclude, we have shown that microscopical structure of traffic ensembles is rapidly changing with the traffic density and is very well described by the one-parametric class of distributions (\ref{p_beta}) where the only parameter is the mental strain coefficient $\beta$ depending on traffic density $\rho.$ In addition, a recent research (Ref. \cite{termodynamical_CA}) shows that a suitable modification (using the knowledge presented in this paper) of Nagel-Schreckenberg cellular model would lead to an interesting progress in the traffic modelling.\\

\begin{figure} \centering
\scalebox{.4}{\includegraphics{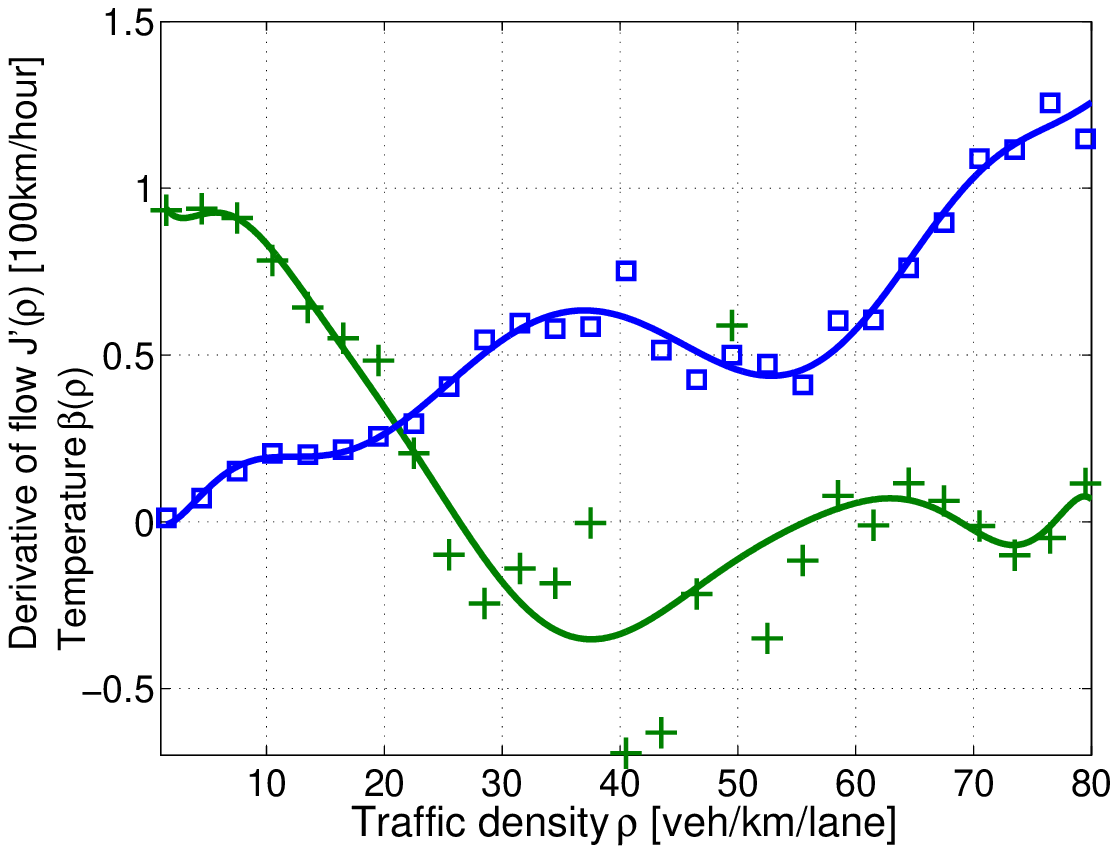}}
\caption{ \small The inverse temperature $\beta(\rho)$ and the first derivative $J'=J'(\rho)$ as functions
of the traffic density $\rho.$ Squares correspond to the inverse
temperature of traffic sample while plus signs display the average
value of $J'=J'(\rho)$ (in kilometers per hour). The
continuous curves represent relevant polynomial
approximations.}
\end{figure}

\emph{Acknowledgements:} We would like to thank Dutch Ministry of
Transport for providing the single-vehicle induction-loop-detector
data. This work was supported by the Ministry of Education, Youth
and Sports of the Czech Republic within the projects LC06002 and
MSM 6840770039.

\end{document}